\documentclass[11pt]{article}
\usepackage{graphicx}
\usepackage{epsfig}
\usepackage{rotating}
\newcommand{\BABARPubYear}    {06}

\newcommand{\BABARConfNumber} {007}
\newcommand{\SLACPubNumber} {12010}

\def\emm{\overrightarrow{m}}
\def\ix{\overrightarrow{x}}
\def\dbarp{\tilde{D}^0}

\def\Dstarz  {\ensuremath{D^{*0}}\xspace}
\def\Dstarm  {\ensuremath{D^{*-}}\xspace}

\def\Kstarz  {\ensuremath{K^{*0}}\xspace}

\def\Kstarp  {\ensuremath{K^{*+}}\xspace}

\def\Kpm      {\ensuremath{K^{\pm}}\xspace}
\def\Dz      {\ensuremath{D^0}\xspace}
\def\piz   {\ensuremath{\pi^{0}}\xspace}
\def\pim   {\ensuremath{\pi^{-}}\xspace}
\def\pip   {\ensuremath{\pi^{+}}\xspace}
\def\KS    {\ensuremath{K_{\scriptscriptstyle S}}\xspace}

\def\mes        {\mbox{$m_{\rm ES}$}\xspace}

\def\to         {\ensuremath{\rightarrow}\xspace}

\def\Dzbpar  {\ensuremath{\Dbar^{0}}\xspace}
\def\rads  {\ensuremath{R_{ADS}}}
\def\rblim {0.185}
\def\rbmeas{\ensuremath{0.091\pm0.059}}
\def\radsval{0.012}
\def\radsstatp{+0.012}
\def\radssysp{+0.010}
\def\radsstatm{-0.010}
\def\radssysm{-0.007}
\def\radssys{0.0076}
\def\radsmeas{\ensuremath{\radsval^{\radsstatp}_{\radsstatm}({\rm 
stat})^{\radssysp}_{\radssysm}({\rm sys}})}
\def\radslim{0.039}
\input pubboard/babarsym

\long\def\inst#1{\par\nobreak\kern 4pt\nobreak
    {\it #1}\par\vskip 10pt plus 3pt minus 3pt}

\begin{document}
{\pagestyle{empty}

\begin{flushright}
\babar-CONF-\BABARPubYear/\BABARConfNumber \\
SLAC-PUB-\SLACPubNumber \\
\end{flushright}

\par\vskip 5cm

\begin{center}
\Large \bf 
Search for {\boldmath $b \rightarrow u$} transitions in
{\boldmath $B^- \to [\Kp\pim\piz]_D K^-$}
\end{center}
\bigskip

\begin{center}
\large The \babar\ Collaboration\\
\mbox{ }\\
\today
\end{center}
\bigskip \bigskip

\begin{abstract}
\noindent
We search for decays of a $B$ meson into a neutral $D$
 meson and a kaon,
with the $D$ meson decaying into $K^+\pi^-\piz$.
This final state
can be reached through the $b \to c$
transition $B^- \to D^{0}K^-$ followed by the doubly
Cabibbo-suppressed $D^0 \to K^+ \pi^-\piz$, or the $b \to u$
transition $B^- \to \Dzb K^-$ followed by the Cabibbo-favored $\Dzb
\to K^+ \pi^-\piz$.
The interference of these two amplitudes
is sensitive to the angle $\gamma$ of the unitarity triangle.
We present preliminary results 
based on 226$\times 10^{6}$ $\epem\to\FourS \to B\Bbar$ events collected with the
\mbox{\slshape B\kern-0.1em{\smaller A}\kern-0.1em
    B\kern-0.1em{\smaller A\kern-0.2em R}}\ detector at SLAC. 
We find no significant evidence
for these decays and 
 we set a limit  $\rads\equiv
   \frac{\Gamma([K^+ \pi^-\piz]_D K^-)+\Gamma([K^- \pi^+\piz]_D K^+)}{\Gamma([K^+ \pi^-\piz]_D K^+)+\Gamma([K^- \pi^+\piz]_D K^-)}
 < \radslim$ at 95\% confidence level,
which we translate with a Bayesian approach into
$r_B \equiv |A(B^- \to \Dzb K^-)/A(B^- \to \Dz K^-)| < \rblim$ at  
95\% confidence level. 

\end{abstract}

\vfill
\begin{center}

Submitted to the 33$^{\rm rd}$ International Conference on High-Energy Physics, ICHEP 06,\\
26 July---2 August 2006, Moscow, Russia.

\end{center}

\vspace{1.0cm}
\begin{center}
{\em Stanford Linear Accelerator Center, Stanford University, 
Stanford, CA 94309} \\ \vspace{0.1cm}\hrule\vspace{0.1cm}
Work supported in part by Department of Energy contract DE-AC03-76SF00515.
\end{center}

\newpage
}

\begin{center}
\small

The \babar\ Collaboration,
\bigskip

%
{B.~Aubert,}
{R.~Barate,}
{M.~Bona,}
{D.~Boutigny,}
{F.~Couderc,}
{Y.~Karyotakis,}
{J.~P.~Lees,}
{V.~Poireau,}
{V.~Tisserand,}
{A.~Zghiche}
\inst{Laboratoire de Physique des Particules, IN2P3/CNRS et Universit\'e de Savoie,
 F-74941 Annecy-Le-Vieux, France }
% ...

\end{center}\newpage

\section{Introduction}
\begin{figure}[hb]
   \begin{center}
      \epsfig{file=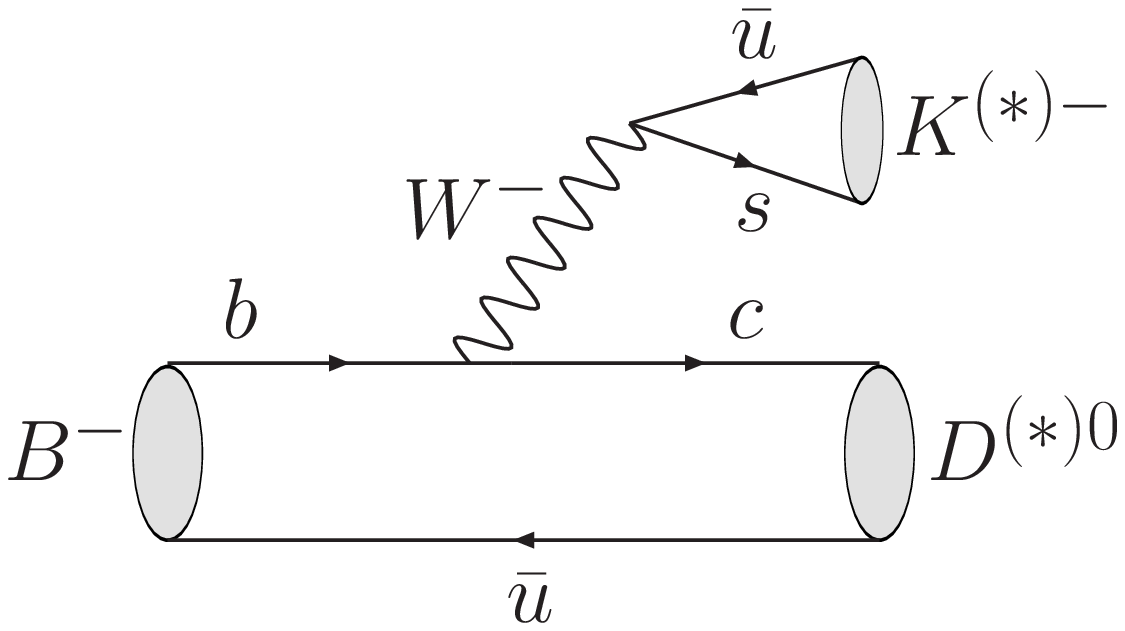,width=0.47\linewidth}
      \epsfig{file=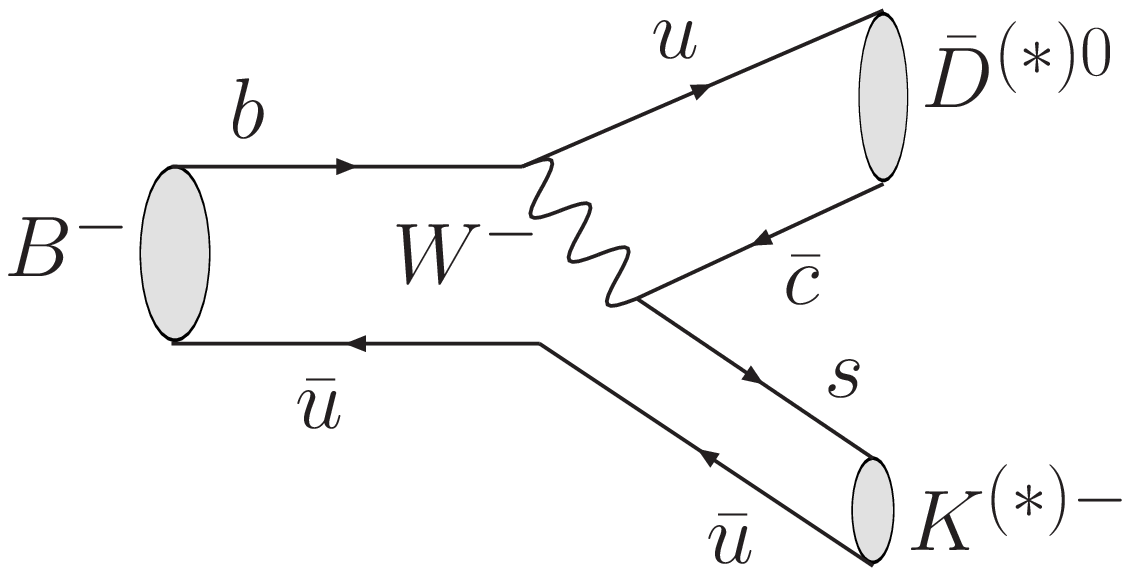,width=0.47\linewidth}
   \caption{Feynman diagrams for the CKM-favored
 $B^- \to D^{0} K^{-}$ and 
the CKM- and color-suppressed $B^-\to \Dzbpar K^{-}$ decays.}
   \label{fig:feynman}
   \end{center}
   \end{figure}

   Following the discovery of \CP violation in $B$ meson 
   decays and the measurement of the angle $\beta$
   of the unitarity triangle~\cite{cpv} associated with
   the Cabibbo-Kobayashi-Maskawa (CKM) quark mixing matrix, focus has turned
   toward the measurements of the other angles $\alpha$ and $\gamma$.
   Following Ref.~\cite{dk1}, several  methods have been proposed to measure the relative weak phase
   between the $B^- \to D^{0}K^{-}$ amplitude, proportional to the CKM matrix element $V_{ub}$ (Fig.~\ref{fig:feynman}),
   and the $B^- \to \Dzbpar K^{-}$ amplitude, proportional to $V_{cb}$. 
This weak phase, which by definition ($\gamma={\rm arg}(-V_{ub}^*V^{}_{ud}/V_{cb}^*V^{}_{cd})$)
is $\gamma$, can be measured from the interference that occurs when the $D^{0}$ and the
   \Dzb\ decay to common final states.

   As an extension of the method proposed in Ref.~\cite{dk2}, we search for 
   $B^- \to [K^+\pi^-\piz]_D K^-$~\cite{cc},
   where the CKM-favored  $B^- \to D^{0} K^-$ decay,
   followed by
   the doubly Cabibbo-suppressed $\Dz \to K^+ \pi^-\piz$ decay,
   interferes with the CKM-suppressed
   $B^- \to \Db^{0} K^-$ decay,
   followed by the Cabibbo-favored $\Dzb \to K^+ \pi^-\piz$ decay.

   In order to reduce the systematic uncertainties, we measure ratios of branching fractions
of the  decay modes of interest in which the two kaons have opposite charge, referred to as 
``wrong sign'' events, to
   the corresponding ones in favored decays, where the kaons have the same charge, referred to as
``right sign'' events. The two ratios we consider, to separate the sensitivity to the suppressed rate and the \CP violation, are:
   \begin{eqnarray}\label{eq:rads}
   \rads &\equiv& 
   \frac{\Gamma([K^+ \pi^-\piz]_D K^-)+\Gamma([K^- \pi^+\piz]_D K^+)}{\Gamma([K^+ \pi^-\piz]_D K^+)+\Gamma([K^- \pi^+\piz]_D K^-)}\\\nonumber
   &=& r_B^{2} + r_{D}^{2} + 2 r_B r_{D} C \cos\gamma\\\nonumber
{A}_{ADS} &\equiv&    \frac{\Gamma([K^+ \pi^-\piz]_D K^-)-\Gamma([K^- \pi^+\piz]_D K^+)}{\Gamma([K^+ \pi^-\piz]_D K^-)+\Gamma([K^- \pi^+\piz]_D K^+)}\\\nonumber
&=& 2r_B r_{D}S\sin\gamma/\rads
   \end{eqnarray}
   \noindent where $D$-mixing effects are neglected, $r_B \equiv \left| \frac{A(B^- \to \Dzb K^-)}{A(B^- \to D^0 K^-)} \right|$, 
$r^2_D \equiv  \frac{\BR(D^0 \to K^+ \pi^-\piz)}{\BR(D^0 \to K^- \pi^+\piz)} $,
  and the $C$ and $S$ parameters take into account the fact that the strong phases of the $D$ decays 
are a function of the decay kinematics. Indicating with $\emm$ a point in the Dalitz plane 
$[m_{K\pi}^2, m_{K\piz}^2]$, with $[{\mathcal{A}}_D(\emm),\delta(\emm)]$  ($[\overline{\mathcal{A}}_D(\emm),\bar{\delta}(\emm)]$) the absolute value and
the strong phase of the $D$ ($\overline{D}$) decay amplitudes,
and with $\delta_B$ the strong phase difference between the two interfering $B$ decays, we have
\begin{eqnarray}\label{eq:cs}
C = \frac{\int \mathcal{A}_{D}(\emm) \overline{\mathcal{A}}_{D}(\emm) \cos(\bar{\delta}(\emm)-\delta(\emm)+\delta_B(\emm)) d\emm}{\sqrt{\int|\overline{\mathcal{A}}_{D}(\emm)|^{2} d\emm \cdot \int|\mathcal{A}_{D}(\emm)|^{2} d\emm }}\label{eqC}
\\
S =  \frac{\int \mathcal{A}_{D}(\emm) \overline{\mathcal{A}}_{D}(\emm) \sin(\bar{\delta}(\emm)-\delta(\emm)+\delta_B(\emm)) d\emm}{\sqrt{\int|\overline{\mathcal{A}}_{D}(\emm)|^{2} d\emm \cdot \int|\mathcal{A}_{D}(\emm)|^{2} d\emm }}\, .
\nonumber\end{eqnarray}

Determining the angle $\gamma$ from the measurements of \rads\ and $A_{ADS}$
 requires extracting  the strong phases, for which the available statistics are insufficient. 
However, the value of $r_B$ determines, in part, the level of interference
   between the diagrams
   of Fig.~\ref{fig:feynman}.  In most techniques for measuring $\gamma$,
   high values of $r_B$ lead to larger interference and better sensitivity to $\gamma$.
   Thus, $r_B$ is a  key quantity for the extraction of $\gamma$ from
other measurements in $B\to D K$ decays~\cite{dalitz}.
In this paper we therefore only measure \rads\ and we constrain $r_B$ by exploiting the fact that in Eq.~\ref{eq:rads} $|C|<1$. 

Both the Belle and \babar\ collaborations have published similar measurements 
but in a different decay chain ($B\to D K$  with 
$D\to K\pi$)~\cite{oldADS}. Unlike those measurements, we can take advantage
of the smaller value of  $r_D$, which is 
$r^2_D=(0.214\pm0.008\pm0.008)\%$~\cite{kpipiz} in 
$D\to K\pi\piz$ decays as opposed to $r^2_D=(0.362\pm 
0.020\pm0.027)\%$~\cite{rdkpi} in 
$D\to K\pi$ decays. This implies that for a given error on \rads, the sensitivity 
to $r_B$ is larger. 

\section{Event Reconstruction and Selection}

   The results presented in this paper are based on 
    an 226$\times 10^{6}$ $\FourS\to\BB$
decays collected 
   between 1999 and 2004 with the \babar\ detector at the \pep2\
   \BF\ at SLAC~\cite{pep2}.  
   In addition, 15.8~fb$^{-1}$ of off-resonance data,
   with center-of-mass (CM) energy 40~\mev below the $\FourS$ resonance,
   is used to study backgrounds from
   continuum events, $e^+ e^- \to q \bar{q}$
   ($q=u,d,s,$ or $c$).
The \babar\ detector is described elsewhere~\cite{babar}. As far as this study is concerned,
 charged-particle tracking is provided by a five-layer silicon
vertex tracker (SVT) and a 40-layer drift chamber (DCH).
In addition to providing precise spatial hits for tracking, the SVT and DCH
also measure the specific ionization ($dE/dx$), which is used for particle
identification of low-momentum charged particles. At higher momenta ($p>0.7$~\gevc)
pions and kaons are identified by Cherenkov radiation detected in a ring-imaging
device (DIRC). The typical separation between pions and kaons varies from 8$\sigma$
at 2~\gevc to 2.5$\sigma$ at 4~\gevc. 
The position and energy of neutral clusters (photons) are
measured with an electromagnetic calorimeter (EMC) consisting of 6580 thallium-doped CsI crystals.
These systems are mounted inside a 1.5-T solenoidal super-conducting magnet.

   The event selection was developed from studies of
   $B\Bbar$ and continuum events simulated with Monte Carlo techniques (MC), and of off-resonance
   data. A large on-resonance data sample of $B^- \to D^0 \pi^-$,
   $D^0 \to K^- \pi^+\piz$ events was used to validate several aspects of the
   simulation and analysis procedure. We refer to this mode
   as $B \to D\pi$.

   In the reconstruction,
 both kaon candidates are required to satisfy kaon identification criteria, which are based on specific ionization loss
measured in the tracking devices and on Cherenkov angles in the DIRC
 and are typically 85\% efficient, depending on momentum and polar angle.
   Misidentification rates are at the two percent level. The \piz\ candidates 
   are reconstructed as pairs of photon candidates in the EMC, each with energy larger than 70\mev
   and lateral shower profile consistent with an electromagnetic deposit, with a total
energy greater than 200\mev, and with
   $118.25 <m_{\gamma\gamma}<145.05 \mevcc$.  
   To account for the correlation between the tails in the distribution of the $K\pi\piz$ invariant mass and the \piz\ candidate mass,
we require the difference between the two measured masses to
 be within 32.5 \mevcc of the expected value of 1729.5\mevcc~\cite{PDG}.
   The remaining background from other $B^\pm \to [h_1 h_2\piz]_D h_3^\pm$~\cite{cc} modes
   is reduced by removing events where any $h_1 h_2\piz$
   candidate, with any particle-type assignment except for the signal hypothesis
   for the $h_1 h_2$ pair, is consistent with a $D^0$ meson decay.

   After these requirements, the background is mostly due to  
   $e^+ e^- \to c \bar{c}$ events, with 
   $\bar{c} \to \Dzb \to K^+ \pi^-\piz$ and $c \to D \to K^-$.
   In order to discriminate against them we use a neural network ($NN$) with
   six quantities that distinguish continuum and $B\Bbar$ events:  
   (i) $L_0 = \sum_i{p_i}$ and (ii) $L_2 = \sum_i{p_i \cos^2\theta_i}$, both
   calculated in the CM frame.  Here, $p_i$ is the momentum and
   $\theta_i$ is the angle with respect to the thrust axis of the $B$ candidate
   of tracks and clusters not used to reconstruct the $B$.
   (iii) The angle in
   the CM frame between the thrust axes of the $B$ 
   and of the detected remainder of the event.
   (iv) The polar angle of the $B$ candidate
   in the CM frame.
   (v) The distance of closest approach between the bachelor
   track and the trajectory of the $D$ meson.  This is
   consistent with zero for signal events, but can be
   larger in $c\bar{c}$ events.
   (vi) the distance along the beams between the reconstructed vertex
   of the $B$ candidate and the vertex of the other tracks in the event. This is consistent
   with zero for continuum events, but is sensitive to the $B$ lifetime for the signal events.

   The $NN$ is trained with simulated continuum and signal events.  
   We find agreement between the distributions of all six variables
   in simulation and in control samples of off-resonance data
   and of $B \to D \pi$ events.  We apply a loose pre-selection  on the $NN$ ($0.4<NN<1.0$) with a 90\% efficiency 
   on signal and a 68\% rejection power over continuum, but then use the $NN$ itself in the likelihood
   fit to fully exploit its discriminant power.

   A $B$ candidate is characterized by the energy-substituted mass
   $\mes \equiv \sqrt{(\frac{s}{2}  + \vec{p}_0\cdot \vec{p}_B)^2/E_0^2 - p_B^2}$
   and energy difference $\Delta E \equiv E_B^*-\frac{1}{2}\sqrt{s}$, 
   where $E$ and $p$ are energy and momentum, the asterisk
   denotes the CM frame, the subscripts $0$ and $B$ refer to the
   initial $\epem$ state and $B$ candidate, respectively, and $s$ is the square
   of the CM energy.  For signal events $\mes$ is centered around the $B$ mass with a 
   resolution of about 2.5 \mevcc, and $\Delta E$ is centered at zero with
an RMS of 0.017 \gev.

   Considering both the right sign and the wrong sign sample, 28621 events 
   survive the selection described above and the loose 
   requirements $|\Delta E|<100\mev$ and $\mes>5.2\gevcc$. The dominant background
   still comes from continuum events, but we also need to take into account
   background from $\FourS\to\BB$ (``\BB'') events. We consider separately the 
   $B \to D\pi$ background since it differs from the signal only in the $\Delta E$ distribution. 
   This decay mode has a very low value of 
 $r_B$  ($\sim r_B(D^0K)\lambda^2$, where $\lambda\sim 0.22$ is the sine of 
the Cabibbo angle), and therefore in the likelihood fit we will consider it as a background only for 
   the right sign sample. 

\section {Likelihood Fit and Results}

   The signal and background yields are extracted by maximizing the extended likelihood
   ${\cal {L}}= e^{-N^\prime}\prod_{i=1}^N{\cal{L}}_i(x_i)/N^\prime!$. Here $N^\prime=N_{DK}+N_{cont}+N_{BB}+N_{D\pi}$ 
   is the sum of the yields of the signal and the three background contributions,
   $\ix=\{NN,\Delta E,\mes\}$, and the likelihood of the individual events (${\cal{L}}_i$) is defined as
   \begin{eqnarray}
\label{eq:pdf}
      {\cal{L}}(\ix)&=&\frac{N_{DK}}{1+\rads}f^{RS}_{DK}(\ix)
      +\frac{N^{RS}_{cont}}{1+R_{cont}}f_{cont}^{RS}(\ix)\\ \nonumber
      &&+\frac{N_{BB}}{1+R_{BB}}f^{RS}_{BB}(\ix)
      +N_{D\pi}f_{D\pi}(\ix)\\ \nonumber
      \mbox{for right sign events and}\\ \nonumber
	{\cal{L}}(\ix)&=&\frac{N_{DK}\rads}{1+R_{ADS}}f^{WS}_{DK}(\ix)
	+\frac{N_{cont}R_{cont}}{1+R_{cont}}f^{WS}_{cont}(\ix)\\ \nonumber
	&&+\frac{N_{BB}R_{BB}}{1+R_{BB}}f^{WS}_{BB}(\ix)\\ \nonumber
	\mbox{for wrong sign events}.
   \end{eqnarray}

We have defined $R$ parameters for the backgrounds with the 
same definition as in Eq.~\ref{eq:rads}. The individual probability density functions ($f$) are derived 
from the MC and 
the three variables are considered uncorrelated in all cases, apart from  $\mes$
 and $\Delta E$ for the $D\pi$ background, since the correlations are not negligible.
 For the latter we have therefore utilized a two dimensional non-parametric 
distribution~\cite{KEYS}. The $NN$ distributions are all modeled with
a histogram with eight bins between 0.4 and 1. The $\mes$ distributions are modeled with a 
Gaussian in the case of the signal, a
threshold function~\cite{argus} in the case of the continuum background, and the sum of a threshold
function and a Gaussian function with an exponential tail in the case of the $\BB$ background. 
Finally, the $\Delta E$ distributions are parametrized with the sum of two Gaussians in the case of 
the signal, an exponential in the case of the continuum background, and a 
double exponential in the case of the $B\overline{B}$ background.  

   \begin{sidewaysfigure}[tb]
      \includegraphics[width=0.3\linewidth]{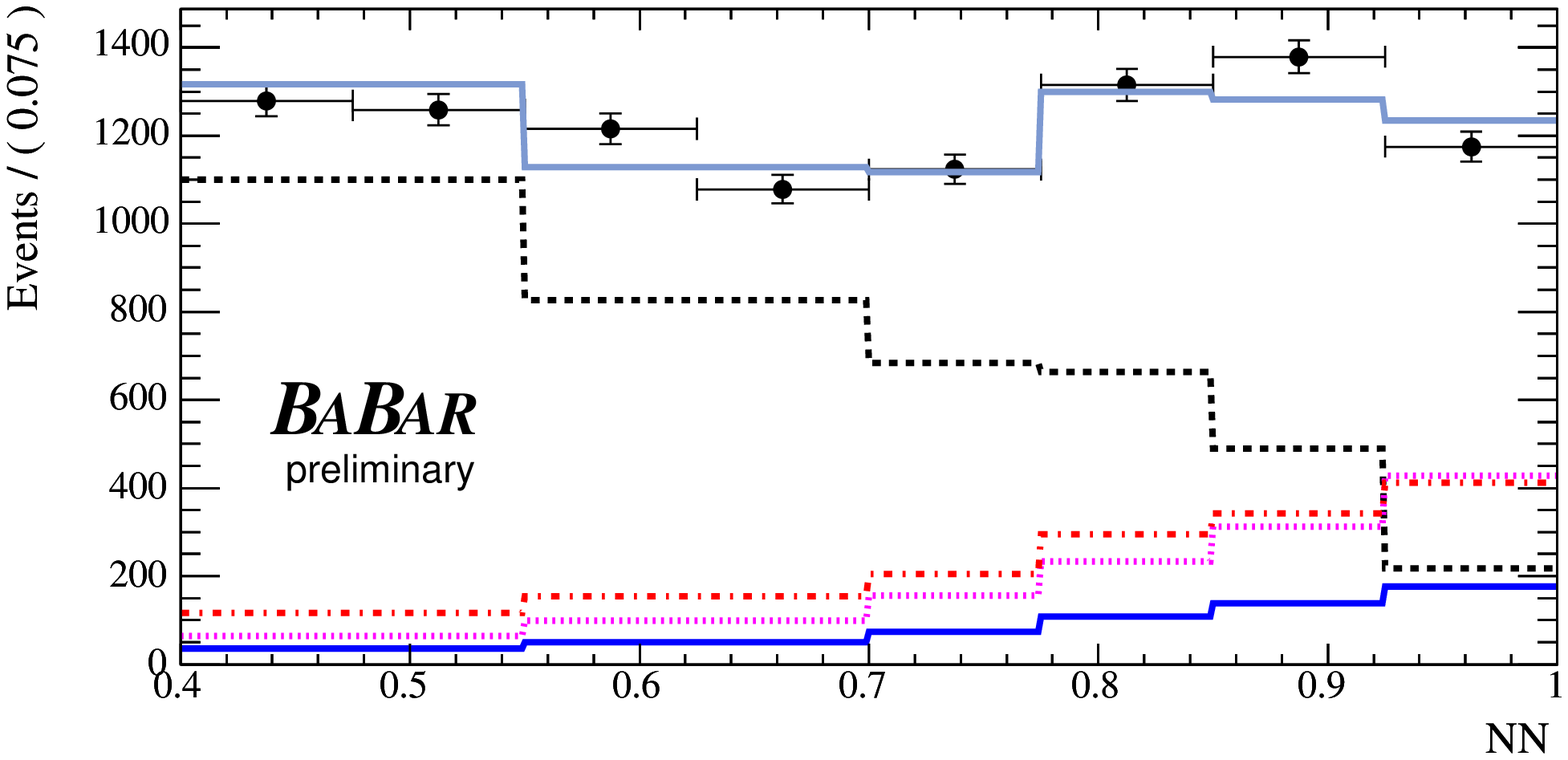}
      \includegraphics[width=0.3\linewidth]{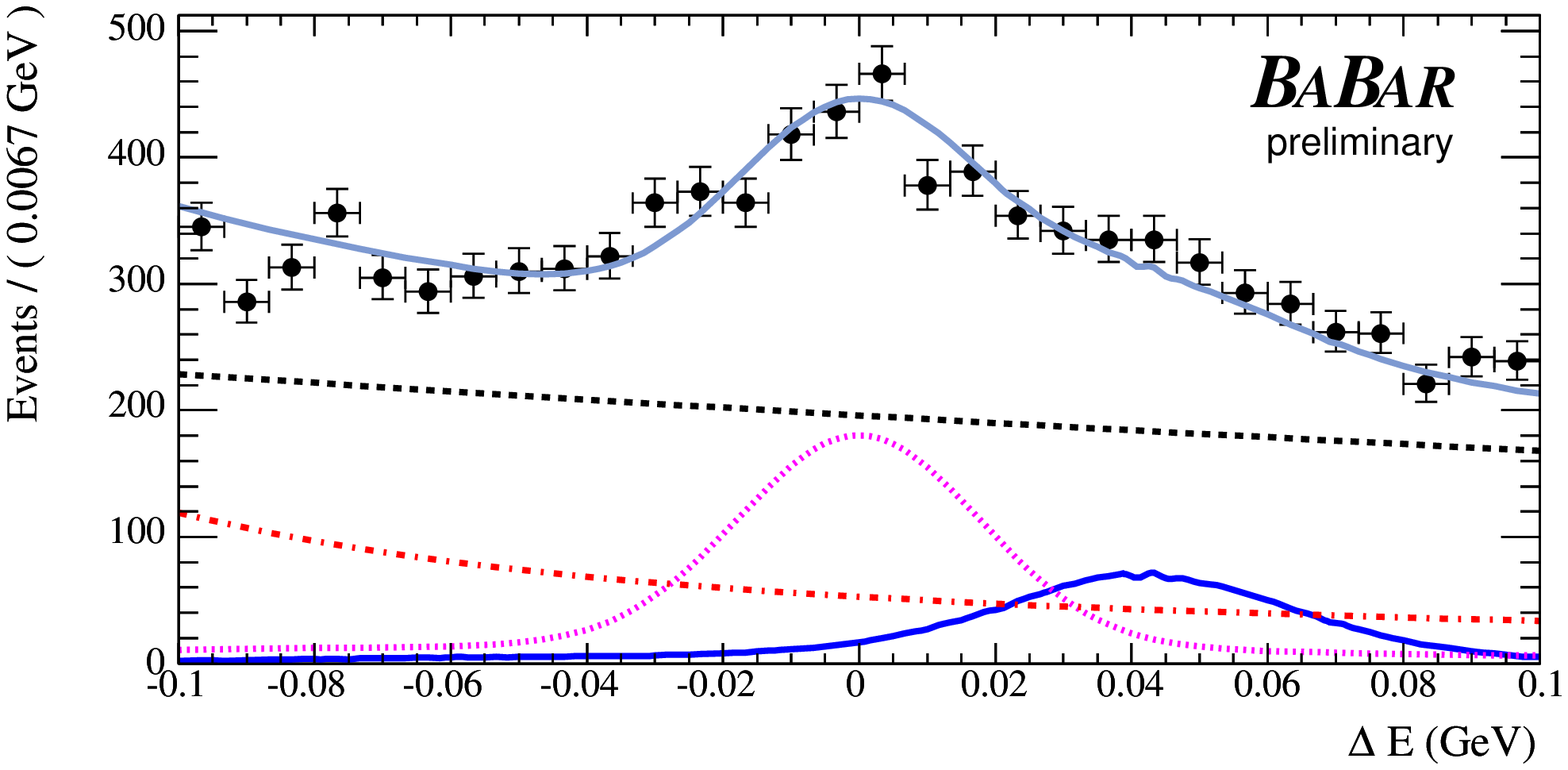}
      \includegraphics[width=0.3\linewidth]{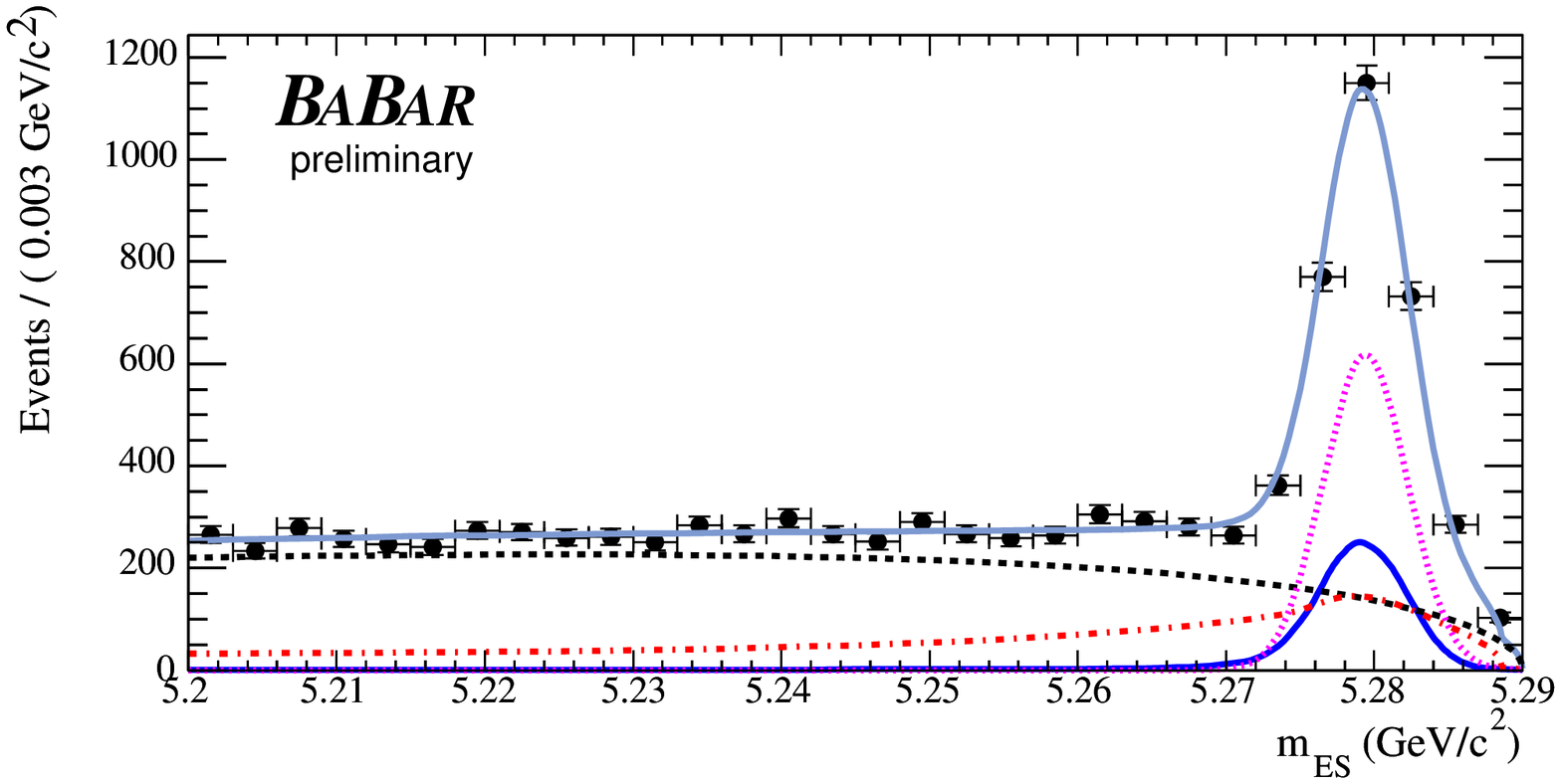} \\
      \includegraphics[width=0.3\linewidth]{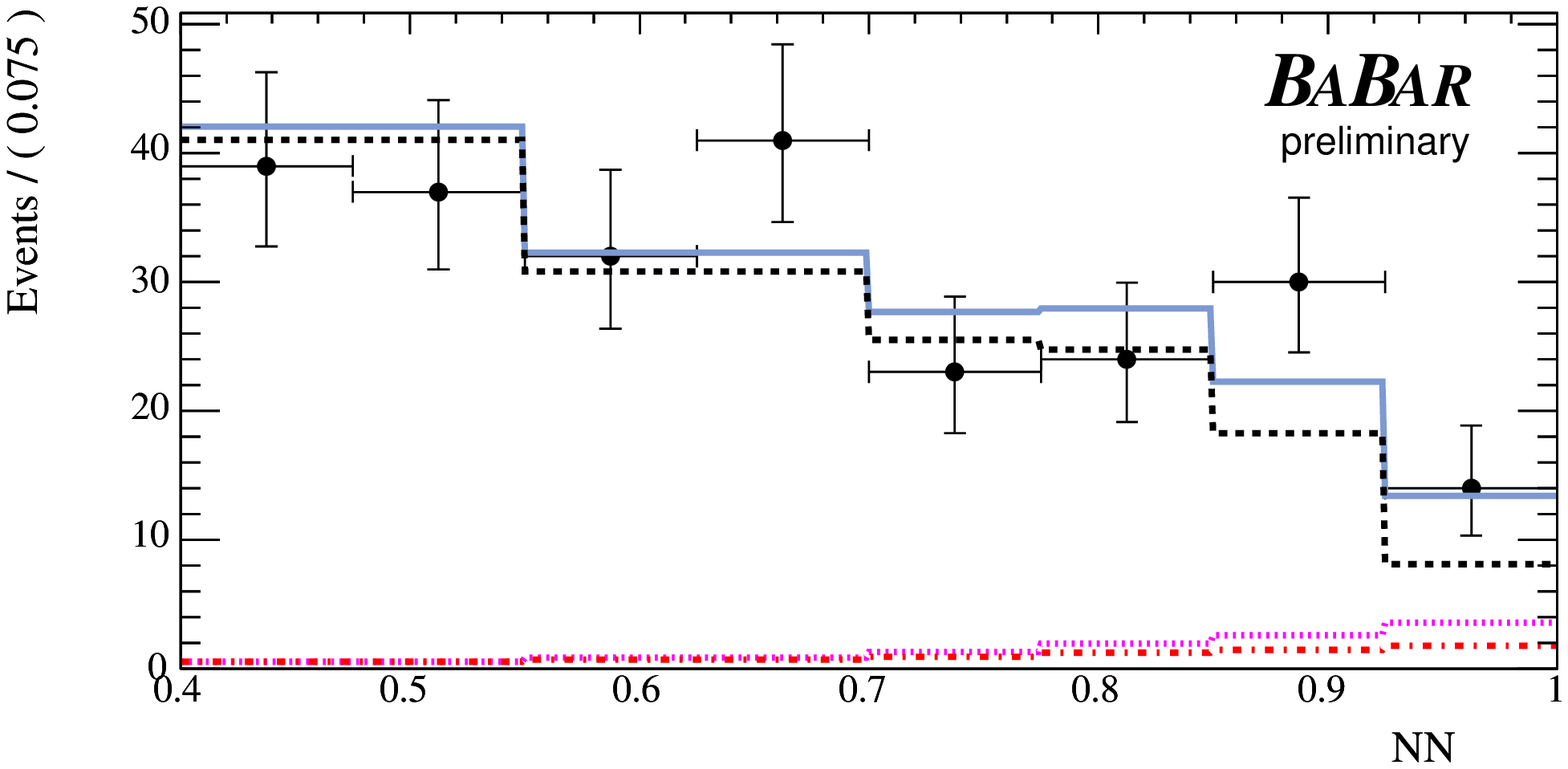}
      \includegraphics[width=0.3\linewidth]{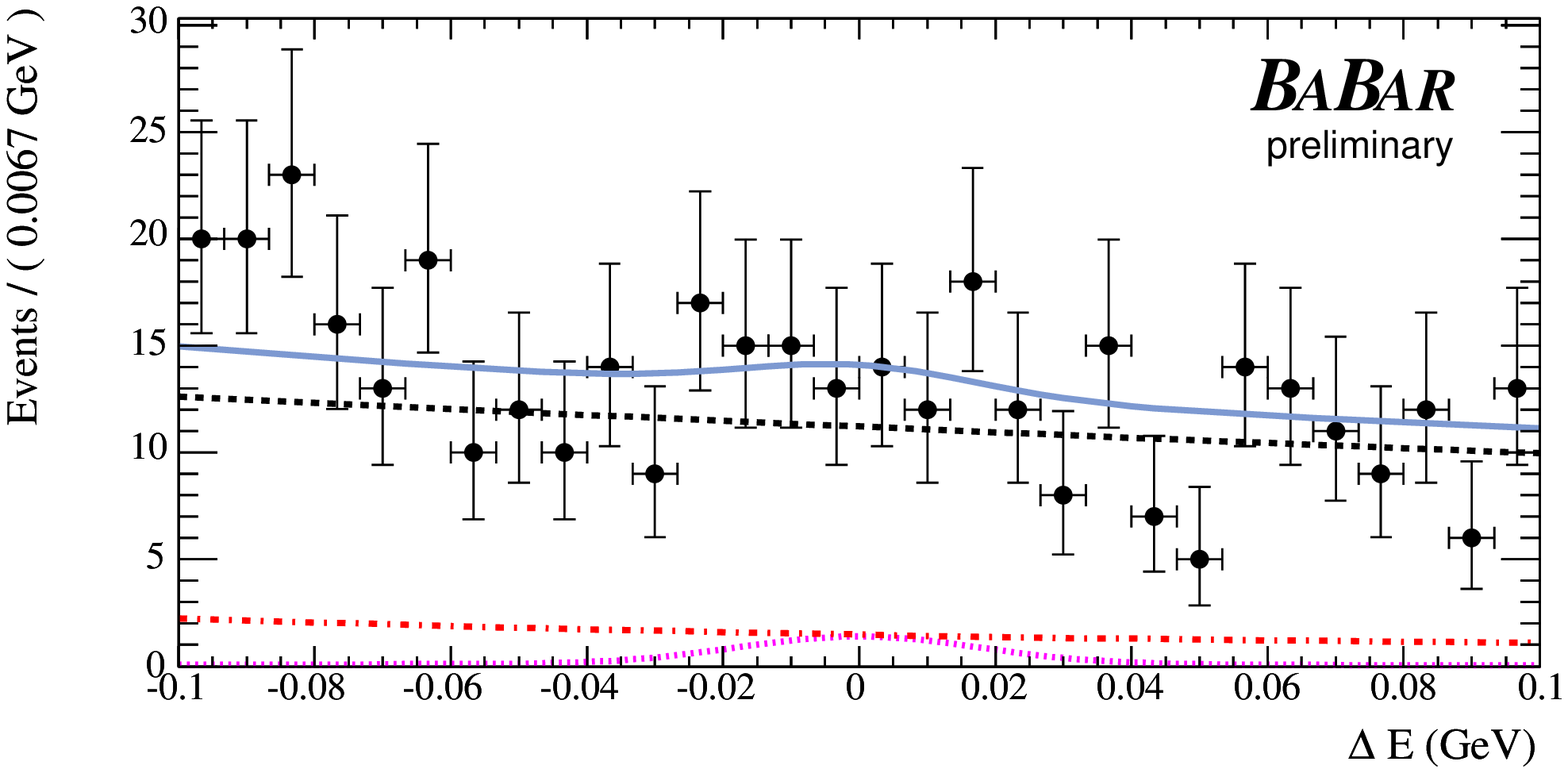}
      \includegraphics[width=0.3\linewidth]{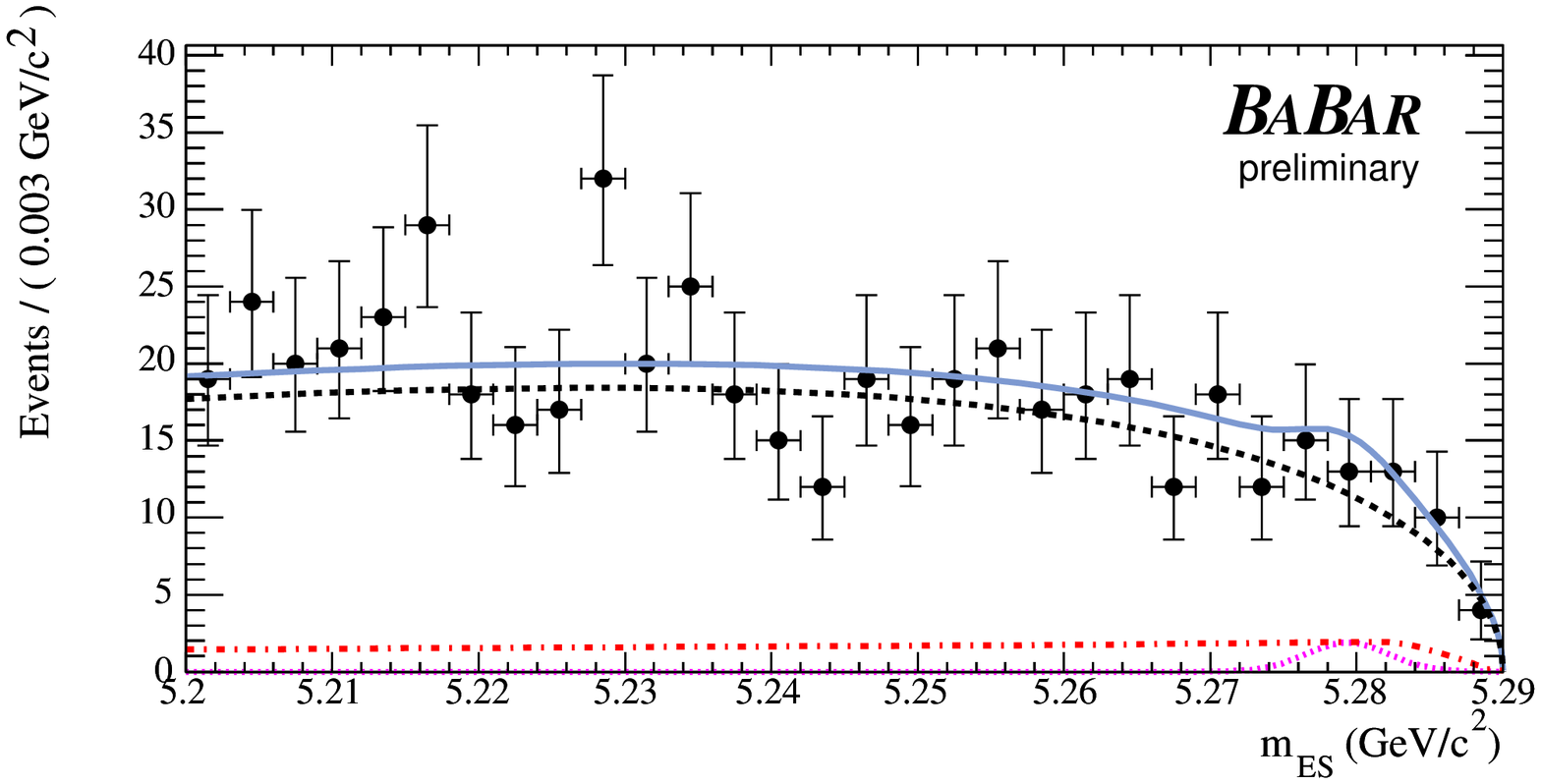} 
   \caption{Likelihood fit projection of the $NN$, $\Delta E$, and $\mes$ distributions separately for the same (top) and opposite (bottom) 
sign sample. To visually enhance the signal, the distributions for the latter sample are shown
after cuts, with a 67\% signal efficiency, on the ratios between the signal and the signal plus background likelihood of all the 
 variables other than the one shown.
  The points with error bars represent the data while
the dashed,  dash-dotted, and solid  lines represent the contribution from continuum, $B\overline{B}$, and $D\pi$ background, respectively.
The dotted line represents the signal contribution, visible only in the same sign sample. } 
   \label{fig:fit}
   \end{sidewaysfigure}

We perform the fit by floating the four total yields ($N_{DK}$, $N_{cont}$, $N_{BB}$, and $N_{D\pi}$), the three $R$ variables 
and the shape parameters of the threshold function used to parametrize $m_{ES}$ for the same and opposite sign continuum 
background.
Figure~\ref{fig:fit} shows the distributions of the three variables in the selected sample (separately for 
same sign and opposite sign events), with the likelihood projections overlaid. 
The fit yields $\rads={\radsval}^{ \radsstatp}_{\radsstatm} $, $N_{DK}=(14.7\pm0.6)\times10^2$, $N_{cont}=(239.3\pm2.1)\times10^2$,
$N_{BB}=(25.5\pm1.6)\times 10^2$, $N_{D\pi}=(6.7\pm0.4)\times10^2$, $R_{cont}=3.05\pm0.07$, $R_{BB}=0.42\pm0.07$.

Equation~\ref{eq:pdf} assumes that the efficiencies are the same between the right and the wrong 
sign signal samples, regardless of the different Dalitz structure. This has been tested on MC and proved
to be true within a statistical error of 4\%. We then consider this as a systematic error on \rads.
We also repeated the fit by varying the probability density function parameters obtained with MC within their
statistical errors and by estimating $f_{cont}^{RS/WS}$ on off-resonance data and $f_{DK}^{RS/WS}$
on exclusively reconstructed $D\pi$ events. To account for the observed variations, we assign a 0.008
systematic error on \rads. The uncertainty due to $B$ decays with distributions similar to the signal, in 
particular $B\to D^{(*)}\pi$, $D^*K$, $D^{(*)}K^*$, and $KK\pi\piz$,
is estimated by varying their  branching fractions within their known errors and found to be 0.00006 on \rads, and therefore negligible.
The absence of further modes which might fake signal has been checked comparing data and MC samples in the sidebands of the $\Delta E$ 
and $m_{D^0}-m_{\pi^0}$ distributions.

Following a Bayesian approach, we extract $r_B$ by defining the a posteriori probability
\begin{equation}
{\cal {L}}(r_B)=\frac{\int p(r_{B},r_D,\xi)
{\cal {L}}(\rads(r_B,r_D,\xi)) d r_D d\xi}{
\int p(r_{D^0 K^+},r_D,\xi){\cal {L}}(\rads(r_{B},r_D,\xi)) dr_D
d\xi dr_B},
\end{equation}
where $\xi=C\cos\gamma$,  $\rads(r_{B},r_D,\xi)$ is given in Eq.~\ref{eq:rads}, and $p(r_B,r_D,\xi)$ 
is the a priori probability for these
three quantities. They are considered uncorrelated, with $\xi$
 and $r_B$  distributed  
flat in the range $[-1,1]$ and $[0,1]$ respectively.
 The a priori probability for $r_{D}$ is a Gaussian
consistent with $r^2_D=(0.214\pm0.008\pm0.008)\%$~\cite{kpipiz}.
 The likelihood ${\cal {L}}(\rads)$ is obtained 
by convolving the  likelihood returned by the fit with a Gaussian of width \radssys, equivalent to the systematic uncertainty. 

\begin{figure}[htb]
     \includegraphics[width=0.4\linewidth]{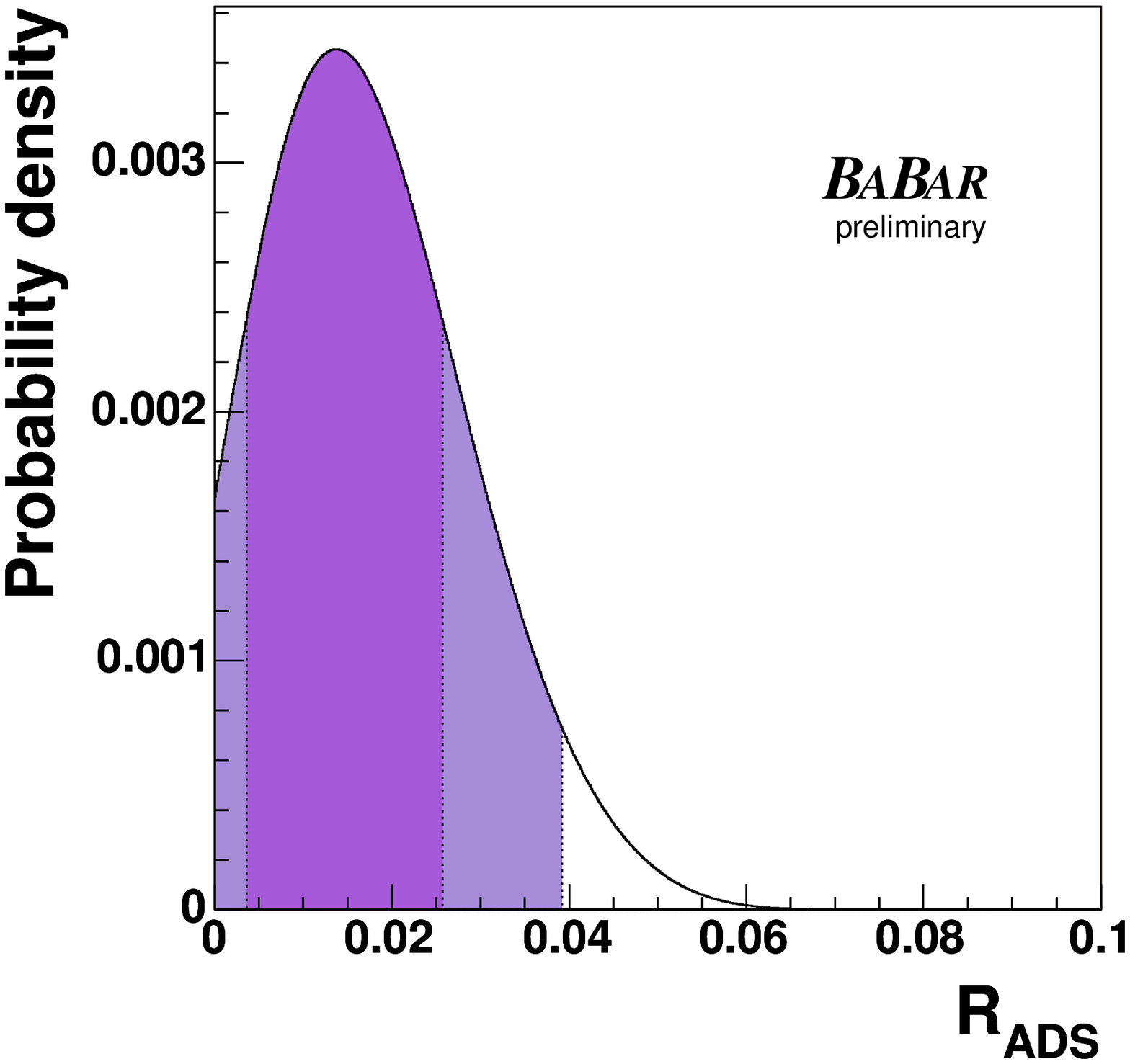}
     \includegraphics[width=0.4\linewidth]{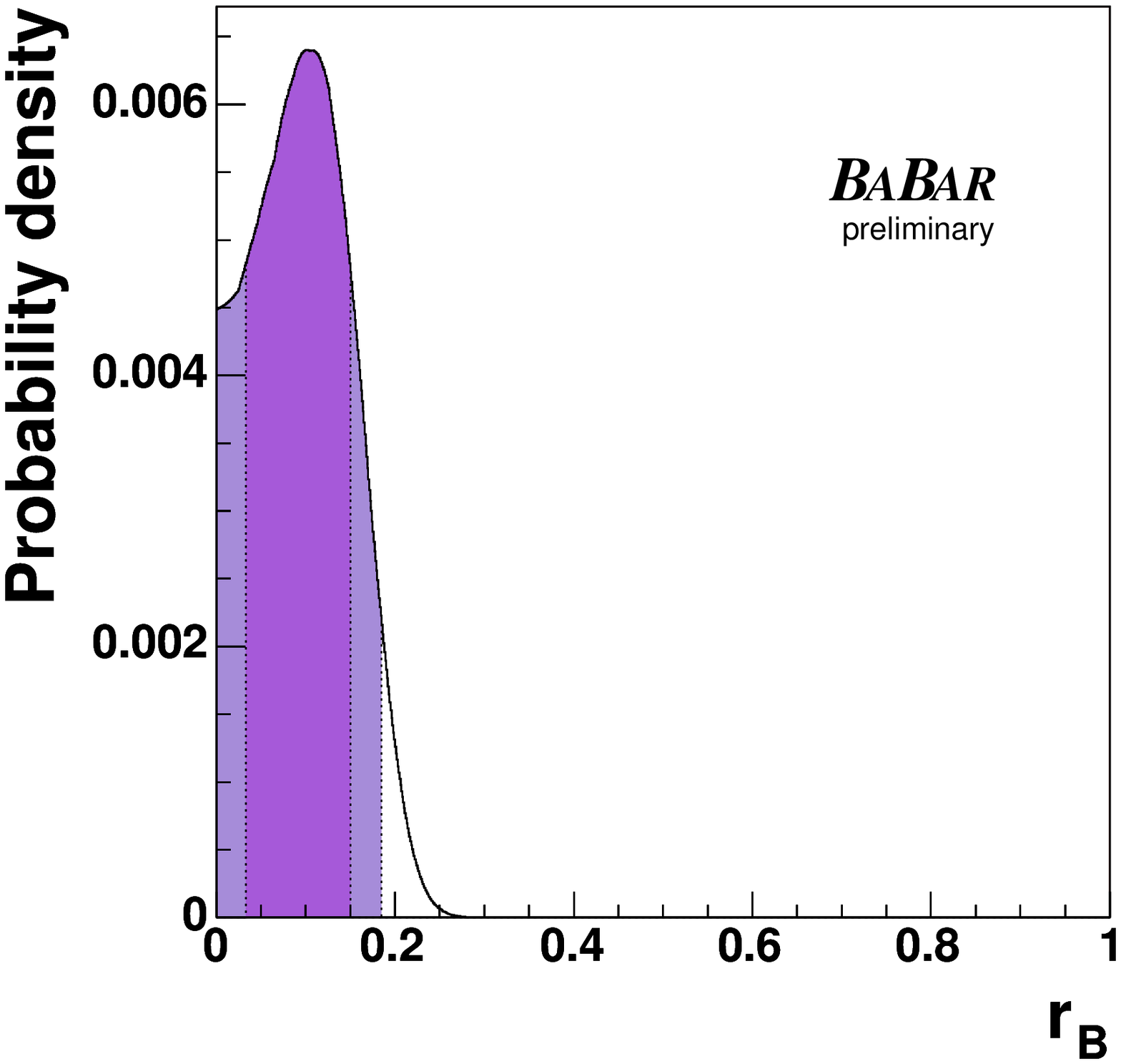}
     \caption{Likelihood as a function of $R_{ADS}$ (left) and of $r_B$ (right).
       While the left plot shows the actual experimental result of the measurements, the right 
       plot is obtained in a bayesian approach assuming flat prior distributions for 
       $r_{B}$,~$C$ and $\gamma$.
       The 68\% and 95\% region are shown in dark and light shading respectively.}
   \label{fig:likelihood_curve}
   \end{figure}

Figure~\ref{fig:likelihood_curve} shows ${\cal {L}}(R_{ADS})$ and ${\cal {L}}(r_B)$. 
We set a 95\% confidence level (C.L.) limit, by integrating  the likelihood  starting from $R_{ADS}=0$ ($r_B=0$), thus excluding unphysical
values, and we define the
68\% C.L. region, for each variable $r=R_{ADS}$ or $r_B$, as the interval where ${\cal {L}}(r)>{\cal {L}}_{min}$ and $68\%=\int_{{\cal {L}}(r)>{\cal {L}}_{min}}
{\cal {L}}(r) dr$. 

\section{Conclusions}

   In summary, we measure the ratio 
   of the rate for the $B^{\pm} \to [K^{\mp}\pi^{\pm}\piz]_D K^{\pm}$  
decay to the favored decay
   $B^{\pm} \to [K^{\pm}\pi^{\mp}\piz]_D K^{\pm}$ to be
  $\rads =\radsmeas$. This result is consistent in central value 
and similar in sensitivity with our completely independent 
previously published result~\cite{oldADS}.
 The measurement is not significant and therefore we set a
   95\% C.L. limit  $\rads < \radslim$. 
   We use this information to infer the ratio of the magnitudes
   of the $B^- \to \Dzb K^-$ and
   $B^- \to D^0 K^-$ amplitudes to be $r_B =\rbmeas$ 
   and consequently set a limit $r_B<\rblim$ at 95\% C.L.

We are grateful for the 
extraordinary contributions of our \pep2\ colleagues in
achieving the excellent luminosity and machine conditions
that have made this work possible.
The success of this project also relies critically on the 
expertise and dedication of the computing organizations that 
support \babar.
The collaborating institutions wish to thank 
SLAC for its support and the kind hospitality extended to them. 
This work is supported by the
US Department of Energy
and National Science Foundation, the
Natural Sciences and Engineering Research Council (Canada),
Institute of High Energy Physics (China), the
Commissariat \`a l'Energie Atomique and
Institut National de Physique Nucl\'eaire et de Physique des Particules
(France), the
Bundesministerium f\"ur Bildung und Forschung and
Deutsche Forschungsgemeinschaft
(Germany), the
Istituto Nazionale di Fisica Nucleare (Italy),
the Foundation for Fundamental Research on Matter (The Netherlands),
the Research Council of Norway, the
Ministry of Science and Technology of the Russian Federation, 
Ministerio de Educaci\'on y Ciencia (Spain), and the
Particle Physics and Astronomy Research Council (United Kingdom). 
Individuals have received support from 
the Marie-Curie IEF program (European Union) and
the A. P. Sloan Foundation.


\begin{thebibliography}{99}
\bibitem{cc}Charge conjugation is implied throughout the paper. 
Also, we use 
    the notation $B^- \to [h_1^+h_2^-\piz]_D h_3^-$ 
   (with $h_i=\pi$ or $K$) for the decay chains $B^- \to \dbarp h_3^-$, 
where $\dbarp$ is either
   a \Dz or a \Dzb and $\dbarp~\to h_1^+ h_2^-$.  
   We also refer to $h_3$ as the bachelor
   $\pi$ or $K$.
\end{thebibliography}
\end{document}